# Contributions of PDM Systems in Organizational Technical Data Management

Zeeshan Ahmed, Detlef Gerhard

Mechanical Engineering Informatics and Virtual Product Development Division (MIVP),
Vienna University of Technology,
Getreidemarkt 9/307 1060 Vienna, Austria
{zeeshan.ahmed, detlef.gerhard}@tuwien.ac.at

*Abstract*— **Product Data Management (PDM) claims of producing desktop and web based systems to maintain the organizational data to increase the quality of products by improving the process of development, business process flows, change management, product structure management, project tracking and resource planning. Moreover PDM helps in reducing the cost and effort required in engineering. This paper discusses PDM desktop and web based system, needed information and important guidelines for PDM system development, functional requirements, basic components in detail and some already implemented PDM Systems. In the end paper investigates and briefly concludes major currently faced challenges to Product Data Management (PDM) community.**

*Index Terms*— **Components, Guidelines, Information, PDM Systems, Current PDM Challenges**

## I. INTRODUCTION

TECNOLOGICAL companies are based on many departments like management, marketing, accounts, production, quality and engineering etc. Every department has its own rules, regulations, data, information, staff and responsibilities which they must have to fulfill. No doubt every department is important and expected to play a vital role in the progress of the company but most of all is the engineering / technical department, which is responsible for the main product's production and rest of all other departments of company are dependent on it. To successfully run the technical department required hardware and software is deployed, processes are initiated and implemented and technical staff is hired.

Till now everything sounds nice, but the problems initiate and start growing as company grows. These problems can happen because of lack in control over engineering processes, rapidly increasing data, lack of presence, lack of coordination among team members (staff), unclear product configurations, loss of experienced staff, conflicts between the central Information Systems (IS) organization, lack of suitable formal communications between departments, bureaucratic and complex engineering change control systems and lack of project and resource management. As the result company can face unnecessary additional increase in costs, delays in product completion, loss in quality and waste of time.

In the past, there were no as such systems available to track and manage implemented process's performance and related product data. This doesn't mean that there was no system for data management; there were some systems based on financial results, corporate plans, manufacturing resource plans, sales data, and personnel information but there was no as such comprehensive system to manage technical data. Engineering data is difficult to manage because of increase in quantity on daily basis, many sources of information like paper, magnetic disks etc. used by many people and computer programs in different functions, has multiple relationships and meanings, exists in many different versions and may need to be maintained for the long time.

## II. PRODUCT / ENGINEERING DATA MANAGEMENT

Product data management (PDM) is computer based system which electronically maintains the organizational data to take advantage in maintaining and improving the quality of product and followed process. PDM is renowned as Engineering Data Management (EDM) and Engineering Document Management Systems (EDMS) because it provides better management and control of engineering data, activities, and changes related to the design and manufacture of product.

Product data management helps in reducing the cost of engineering, reducing effort in product development life cycle, reducing time in engineering change handling and new product development, improving the quality and services of product by making a strong effect on competitively, market share and revenues [1]. Moreover PDM is also capable of handling the business process flows, change management, revision control, product configurations, product structure management, project tracking, resource planning, product variations and versions [2].

Major objectives of product data management are

1) Deliver and support the products at the time
2) Improve product quality
3) Improve team co-ordination
4) Increase customization of products
5) Maintain product configuration based information
6) Manage the large volumes of data generated by computer-based systems
7) Reduce engineering environment based problems
8) Provide better access to information
9) Provide better reuse of design information
10) Solve some currently existing problems
11) Provide common data ware housing
12) Secure engineering data's originality
13) Prevent error creation and propagation
14) Maintain engineering data in reusable form
15) Efficiently manage numeric data, text information, graphic information and increasingly voice data.

### III. PDM SYSTEMS

In early 1970's and 1980's there was no as such systems to automate data management and then years ago Computer Integrated Manufacturing (CIM) was introduced but not seemed to be successful in product data management. PDM System are introduced and used to manage engineering data, activities and process through better control of engineering data, activities, changes and product configurations. PDM provide a backbone for the controlled flow of engineering information throughout product life cycle by using engineering data, such as CAD, ERP and field service. More over PDM also support product teams and techniques by providing Concurrent Engineering in improving engineering workflow and PDM systems address issues such as control, quality, reuse, security and availability of engineering data.

Currently, to integrate and manage all applications, information, and processes during the associated product life cycle five main user functions .i.e., data vault and document management, workflow and process management, product structure management, parts management, and program management are provided by PDM systems including Metaphase (SDRC), SherpaWorks (Inso), Enovia (IBM), CMS (WTC), Windchill (PTC), and Smarteam (Smart Solutions) [6].

Now a days PDM Systems are divided in to two kinds .i.e., Data tier and Middle tier. Data tier PDM Systems contains database and file vaults where as the Middle tier PDM Systems contains business logic and connection mechanisms. PDM Systems are available in both desktop and web form but there are several advantages of using Web based PDM Systems over desktop PDM systems .i.e., capable of providing mechanism for worldwide electronic information storing, sharing, reshaping and offers global business centralized communication and platform independency.

*A. PDM System Components*

Every PDM System consists of some basic components which are most of the time need to be developed and improved, which are

1) Data warehouse to store engineering information.
2) Information management module controls to access, store, integrate, secure, recover and manage information in data warehouse.
3) Basic networked computer environment based infrastructure.
4) Interface module to support user queries, menu and form driven inputs and report.
5) Information and Workflow Structure Definition Modules to manage resources, events and responsibilities.
6) Information Structure Management Module to produce exact structure based on maintained information in the system.
7) Workflow Structure Definition Flow and content of engineering activities.
8) Workflow Control Module to control and coordinates engineering processes.
9) System Administration Module to set up administration and maintain the configuration of the system.

*B. PDM System Development Guidelines / Steps*

To develop a PDM System consisting of at least above mentioned basic components in section 3.1 number guidelines/steps are need to be followed, which are

1) Before starting the PDM System try to understand the current engineering process, environment, business objectives, the user requirements and the functionalities of the currently deployed systems in targeted company.
2) Develop information model to describe the current situation of company.
3) Involve top management, the project team leader, and the other project team members in PDM System design development process.
4) Interview the staff of company in small groups of 2, 3 or 4 persons involved many activities, and collect details of the various resources they use.
5) Read company's explicitly provided information and review relevant documents and systems.
6) Collect a lot of information about the engineering data, how it is created and used in the company.
7) Try to understand data standards and data ownership of the company.
8) Use top down approach or bottom up approach in defining boundaries of collected contents.
9) Analyze all the phases of product life cycle used in company like process planning, conceptual detailed design, analysis, production planning, engineering, quality control, drafting, tooling, numerical control (NC) programming, machining, assembling, identifying for each activity the input, use, creation, storage, output of information, packaging and distribution.

10) Address almost all individual activities like engineering design, process planning in detail.

11) Define deliverables at the end of each activity including description, overall activity process flow diagram, data flow diagram (DFD), descriptions of associated documents and systems used in each activity.

12) Prepare a list of identified problems, proposed improvements and proposal for the next phase of PDM activity.

13) Describe each activity, its objective and its position in the overall activity flow. If necessary then break down the activity into its constituent tasks.

14) Try to understand how packets of information are created, modified and moved between activities

15) While examining all activities individually understand the information and involved parameters concerning the volume of information.

16) Identify the number and description of persons involved in the activities.

17) Identify most important factors and business metrics for management to reduce lead times and improve product quality.

18) Quantify the obtained information from management to have more meanings and measure of progress.

19) Analyse business objectives to have a clear business focus and to prevent from drowning in the sea of information.

20) Produce detailed workflow diagrams to understand the workflow.

21) Try to understand the outlined picture of company from engineering the point of view to have knowledge about the number of employees involved, their location, functions and the way they store and communicate information with each other.

22) Identify typical milestones, events, and management activities by looking into the structure of product.

23) Understand data security and integrity issues.

24) Understand review, release and change processes.

25) Identify information repositories containing engineering drawings, CAD/CAM, electronic data, and alphanumeric engineering documents.

26) Draw an inventory of existing information system related to engineering information consisting of CAD, CAM and CASE.

27) Quantify the Information Systems (IS) abilities and resources to avoid proposing a solution that requires IS support that the company's IS functions are incapable of supplying.

28) Try to identify illogical, wasteful and unnecessary or duplicate activities.

29) When all the required information is obtained then start analysis, cross check and structuring information.

30) Organize and manage resultant information consisting of overall activity flow diagram, description of each of the major activities, flow diagrams for each major activity, high level information flow diagram, data flow diagram (DFD) for each individual activity, descriptions of documents and systems used in each activity, list of problems identified, list of proposed improvements, proposal for the next phase of EDM/PDM activity, list of people interviewed to present to management of company.

*C. PDM System Basic Functional Requirements*

Some basic functional requirements are required to be implemented in a good PDM System are

1) The PDM system should be flexible, because every organization has different implementation and priorities.

2) The PDM system should have friendly, well designed and organized graphical user interface (GUI).

3) The PDM system should consist of open architecture which can integrate with already used tools and technologies in organization.

4) The PDM system must be platform independent and function across heterogeneous networks to assure a common and concurrent engineering environment.

5) The PDM system should be rule based and event driven to control the system functions.

6) Ensure PDM System Design release management, Change management, Product Structure Management, Classification and Program Management by

7) Providing secure access control, establish data relationships, check in and checkout, global release definitions, user lists and meta data management.

8) Specifying process definitions, in other words "Who will approve what, why and when".

9) Providing part list and bill of material functions, part definitions, part relationship attributes and the ability to associate product defining art with parts and structures.

10) Providing tools to search and retrieve standard parts and existing designed and stored data.

11) Creating work breakdown structures and schedules resources.

## IV. VALUED METHODOLOGIES TOWARD PDM SYSTEMS

In this section we briefly address some already proposed and implemented methodologies providing values to product data management applications.

*A. Partial Distance Map (PDM)*

Authors have proposed an interactive high dimensional indexing scheme based on Partial Distance Map (PDM) [4] to speed up the process of information retrieval especially from Chinese calligraphic character database. According to the authors, due to the high level complexity of Chinese language characters no as such efficient technique exists to retrieve and index large Chinese calligraphic character databases.

Authors have used Approximate Point Context (APC) to search characters and proposed PDM to established links between semantic level concepts of characters and low level shaped features through user relevance feedback. Moreover during information retrieval process Pruning Distance Table (PDT) is used and dynamically updated. Authors developed a pseudo search algorithm based on Hyper Center Reallocation (HCR), an Approximate Minimal Bounding Hypersphere and Clustering based method, used to find out First Nearest Neighbor (1-NN). The pseudo algorithm works in two steps, first it calculates 1-NN using HCR based on submitted query, then searches with a small radius and increase the size of radius step by step to form a bigger query sphere iteratively. Once the number of candidate characters is become larger than k, the (|S|−k) characters which are farthest to the query one are identified and removed from S. In this way, just the k nearest neighbor characters is returned.

Proposed techniques is also evaluated by authors in an extensive performance study, which was resulted with the information that PDM is superior to iDistance and NB-tree in terms of both I/O and CPU cost.

### B. Component based Product Data Management System

Authors have proposed adequately available, secure, stable, and reliable platform independent Web based PDM System called Component based Product Data Management System (CPDM) [5] to take advantage in decreasing the cycle time for introducing new technology and products. The architecture of CPDM is based on Component Based Development (CBD), implemented using J2EE and consist of consists of three tiers .i.e., Presentation tier, Business logic tier and Data tier. Presentation tier is designed to access the system through a Web browser, Business logic tier is implemented to handle the core PDM functionality and Data tier is composed of a database and vault for the physical files.

CPDM is developed using Rational Unified Process (RUP) in a next generation military system for a local military company. Developed CPDM is successfully deployed and several advantages in terms of productivity and competitiveness .i.e., reducing design hours, eliminating reworking due to design inconsistency, reducing time to market, and reducing parts inventory are obtained.

### C. Web-based product data management (WPDM)

Authors have proposed an easily extendable Web based product data management (WPDM) [6] to provide product database management through internet, access to users at different locations, especially those on different networks, optimize product cycle time and platform independency.

The architecture of WPDM is based on three-tier client/server system .i.e., Back end database, Front end design and Middle layer, allow users to create, view and manipulate product data through the World Wide Web. Back end database is a data vault used to store and manage product attribute data, documentary information and relationships between data. Front end design uses a web browser based front end to access WPDM. Middle layer is a centralized control between back end database and front end user. The designed architecture of WPDM is implemented using Java Servlets, JSP and JDBC and capable of handling user input and managing data in database.

## V. PDM CHALLENGES

Now days PDM community is facing some major challenges, which are

1) Successful implementation of a PDM based application in organization (especially large) because it is time consuming, expensive and most of the staff belongs to corporate management, top level management, engineering management and other engineering and IT professionals do not give importance to PDM and without these person's support it is quite difficult to implement it [2].

2) Time to market benefits of concurrent engineering while maintaining the control of data and distributing it automatically to the respective persons [3].

3) Optimized and decreased cycle time for introducing new technology and products.

4) PDM System deployment.

5) People don't want to use PDM because of several reasons .i.e., don't want to involve in low level technical and business issues, don't want to spend money, look for fast payback projects, don't have time, too much inertia in this company, lack of trust of users on management, job insecurity, incapable of handling PDM systems, CAD, PDM is immature, not flexible, its risky, not intelligent to get the right information, limited in memory, electronic and dependent on system, slow.

6) PDM Systems are required to be platform independent because in the new business model, it is nearly impossible to mandate that all the potential users choose the same platform or the same operation system.

7) PDM Systems are required to be easily extendable because whenever new features are demanded users must reinstall or upgrade the client application completely.

8) Traditional PDM Systems are not adequately available, secure, reliable, and scalable for global enterprise services.

9) Standardized Web based frameworks for Web based PDM systems development.

10) Homogeneous network configured PDM, provide access to users at different locations, especially those on different networks.

## VI. CONCLUSION

In this paper we have discussed Product Data Management (PDM) as the major role player in supporting organizational data management to take advantage in increasing the quality of products with in less resources and cost. Focusing on PDM Systems we have briefly discussed important guidelines and required information for system development, basic components and some already implemented PDM Systems. No doubt PDM desktop and web based systems have contributed alot in improving the engineering processes, business process flows, change and product structure management, project tracking and resource planning but still there are some challenges which need to be talked.